\documentclass[twocolumn,a4paper,amsmath,amssymb]{revtex4}
\usepackage{graphicx}
\usepackage{times}
\usepackage{amsmath}
\usepackage{braket}
\usepackage{xcolor}
\usepackage{natbib}
\usepackage{dsfont}
\usepackage{hyperref}
\usepackage{pdfpages}

\renewcommand{\eqref}[1]{Eq.~(\ref{#1})}

\newcommand{\figref}[1]{Fig.~\ref{#1}}
\newcommand{\removedD}[1]{{\color{gray}{#1}}}
\renewcommand{\removedD}[1]{{}} %comment this line to show removed parts

\newcommand{\CEE}[1]{{\color{black}{#1}}}
\newcommand{\CEEE}[1]{{\color{black}{#1}}}

\renewcommand{\eqref}[1]{Eq.~(\ref{#1})}

\newcommand{\appref}[1]{\hyperref[#1]{Appendix~\ref*{#1}}}
\newcommand{\tabref}[1]{\hyperref[#1]{Table~\ref*{#1}}}

\begin{document}
\title{Quantum limited amplification and entanglement in coupled nonlinear resonators}
\author{C. Eichler$^1$, Y. Salathe$^1$, J. Mlynek$^1$, S. Schmidt$^2$, A. Wallraff$^1$}
\affiliation{$^1$Department of Physics, ETH Z\"urich, CH-8093, Z\"urich, Switzerland.}
\affiliation{$^2$Institute for Theoretical Physics, ETH Z\"urich, CH-8093, Z\"urich, Switzerland.}
\date{\today}
\begin{abstract}
We demonstrate a coupled cavity realization of a Bose Hubbard dimer to achieve quantum limited amplification and to generate frequency entangled microwave fields with squeezing parameters %\cite{Castellanos2008,Eichler2011a,Flurin2012,Menzel2012}
well below -12 dB. In contrast to previous implementations of parametric amplifiers our dimer can be operated both as a degenerate
%\cite{Castellanos2008}
and as a nondegenerate amplifier.
%\cite{Bergeal2010a}.
The large measured gain-bandwidth product of more than 250$\,$MHz for nondegenerate operation and the saturation at input photon numbers as high as 2000 per $\mu$s are both expected to be improvable even further, while maintaining wide frequency tunability of about 2$\,$GHz. Featuring flexible control over all relevant system parameters, the presented Bose-Hubbard dimer based on lumped element circuits has significant potential as an elementary cell in nonlinear cavity arrays for quantum simulation.
\end{abstract}
\maketitle
%\Cite{Karabalin2009,Giampaolo2009,Habraken2012,Schmidt2010a}
%
%\section{Input-Output relations for the Parametric amplifier}
{The high level of control achievable over collections of massive or massless particles, such as atoms, spins or photons, enables the detailed study of intricate many-body phenomena in man-made quantum systems \cite{Houck2012}. In this context coupled nonlinear resonators
%embedding a nonlinear element
both provide a viable avenue for studying light-matter interactions and constitute a generic building block for photonic quantum simulators of strongly interacting systems \cite{Greentree2006,Angelakis2007,Hartmann2006}.
%Hartmann2006,Carusotto2013,Schmidt2013}.
Therefore, their theoretical and experimental investigation is pursued in a wide variety of physical settings such as photonic structures \cite{Gerace2009,Abbarchi2013}, optomechanical systems \cite{Ludwig2012,Palomaki2013,Okamoto2013} and superconducting  circuits \cite{Zant1992,Schmidt2010a,Raftery2013}.
}

The remarkable progress in quantum science using microwave radiation
%\cite{Devoret2013}
has stimulated broad interest in low noise amplification \cite{Caves1982,Yurke1988} and has lead to the development of novel versatile amplifiers in the recent years \cite{Castellanos2008,Tholen2007,Yamamoto2008,Kinion2008,Bergeal2010,Hover2012,Mutus2013,Abdo2013}. Many of these implementations rely on parametric processes in which the noise temperature of the amplifier is solely limited by the radiation temperature of the input fields, ultimately by the vacuum fluctuations \cite{Clerk2010}. In parametric amplification the presence of a signal stimulates conversion processes from a pump field into the signal field, while creating an additional idler field. When signal and idler field occupy the same mode, this is referred to as degenerate parametric amplification, whereas in nondegenerate amplifiers the signal and idler modes are separated either spatially or in frequency \cite{Clerk2010}.
%Either of these amplification schemes can be superior depending on the desired application.
While degenerate parametric amplifiers \cite{Castellanos2008} are often preferable for the fast dispersive readout of qubits
%\cite{Vijay2011}
in quantum feedback protocols,
%\cite{Riste2012a},
%where information is typically encoded in a single quadrature of the field,
nondegenerate amplification \cite{Bergeal2010} can be more practical for multiplexed readout, the measurement of photon correlation functions %\cite{Bozyigit2011}
and more general applications in which amplification is to be independent of the phase of the signal relative to the pump.

\begin{figure}[b]
\centering
\includegraphics[scale=1]{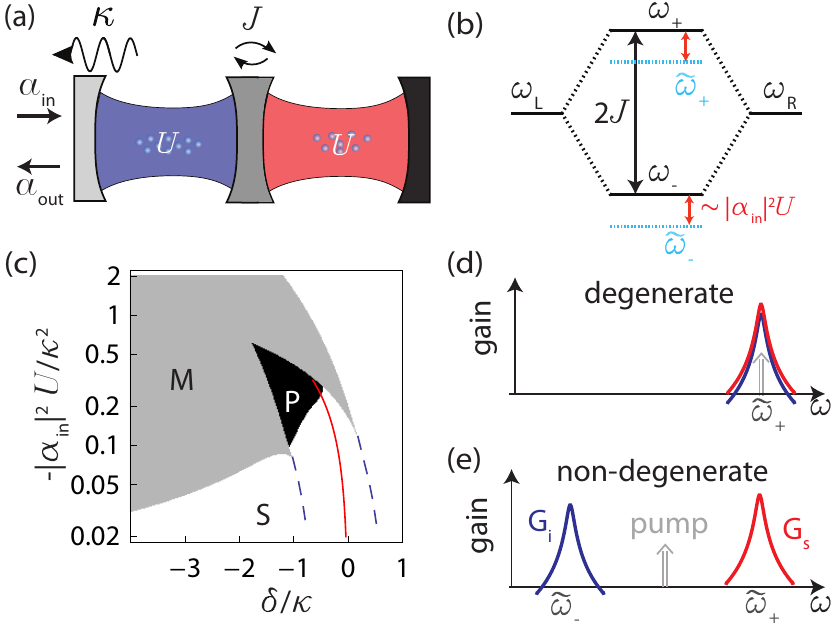}
\caption{(a) Optical frequency representation of the Bose-Hubbard dimer, illustrated as two cavities each with on-site interaction strength $U$ and  coupled with hopping rate $J$. The left cavity emits into a transmission line at rate $\kappa$. (b) Mode structure of the dimer and drive induced red-shifts. See text for details. (c) Calculated phase diagram of the dimer driven with a coherent input field $\alpha_{\rm in}$ at detuning $\delta$ from the bare resonance frequency $\omega_0$ for $J=0.7\kappa$ and  $U<0$.
%With increasing drive rate the system goes from a stable (S) either into a parametrically instable phase (P) or into a multi-stable phase (M).
%depending on the drive detuning $\delta$.
The red line indicates drive configurations with vanishing field amplitude in the left cavity. The dashed blue lines indicate redshifted frequencies $\tilde{\omega}_-$ and $\tilde{\omega}_+$. (d) Driving the system at frequency $\tilde{\omega}_+$ (gray vertical arrow) results in degenerate parametric amplification, with signal gain $G_s$ (red) and idler gain $G_i$ (blue) both occupying the symmetric mode. (e) Signal and idler gain become nondegenerate when driving the system in between $\tilde{\omega}_+$ and $\tilde{\omega}_-$.
}
\label{fig:JPD1}
\end{figure}

\CEEE{Here, we consider a system described by two bosonic modes
%described by field operators
$a_L$ and $a_R$, which are coupled with hopping rate $J$ and have an on-site interaction strength $U=U_L=U_R$, see generic representation in \figref{fig:JPD1}(a).} In a frame rotating at the bare cavity frequency $\omega_0=\omega_{L}=\omega_R$ the system is described by the Bose-Hubbard-dimer Hamitonian \cite{Sarchi2008} $$H/\hbar=J(a_L a_R^\dagger + a_R a_L^\dagger) + \frac{U}{2} ((a_L^\dagger)^2  a_L^2 + (a_R^\dagger)^2 a_R^2).$$
%with photon number $n_i \equiv a^\dagger_i a_i$.
%~\cite{Eichler2014aSupp}.
While the left mode ($L$) is coupled with rate $\kappa$ to a transmission line the right mode ($R$) is only coupled to the left mode.
%Briefly discuss the different parameter regimes which has been addressed and what it has been used for, so far.
We consider the parameter regime $|U|\ll J\lesssim \kappa$ to achieve quantum limited amplification. We note that the design presented below allows us to realize circuits in which any of the three rates may dominate over the other two. Due to the hopping term $J$ the left and right modes hybridize and form symmetric and antisymmetric eigenmodes  $a_{+}$ and $a_{-}$.  The corresponding eigenfrequencies are split by $2J = \omega_{+} -\omega_{-}$ around the bare cavity frequency $\omega_0$. A coherent drive field  $\alpha_{\rm in}$ applied to the dimer in combination with the nonlinearity $U$ shifts the effective resonance frequencies to $\tilde{\omega}_-$ and $\tilde{\omega}_+$, see \figref{fig:JPD1}(b). The effect of finite detuning $\omega_L-\omega_R$ between the left and right mode and unequal interaction strengths $U_L\neq U_R$ are considered in the supplementary material.

We have calculated the phase diagram of the Bose-Hubbard dimer when driven coherently at a detuning $\delta$ from the bare cavity frequency $\omega_0$ using a semiclassical approximation. For drive rates $|\alpha_{\rm in}|^2$ on the order of $\kappa^2/U$ the system undergoes a sharp transition
%from a stable regime (S) into either a multi-stable  region  (M) or a region with unique but parametrically instable steady-state solution (P).
%%As illustrated in \figref{fig:JPD1}(b), this phase transition occurs when the drive induced nonlinearity $U |\alpha_{in}|^2/\kappa$ is on the order of $\kappa$.
%While the multi-stable region is characterized by the appearance of multiple solutions to the classical equation of motion, we identify the parametrically instable region from the existence of a positive eigenvalue in the stability matrix
from a regime with one stable solution (S) into either a multi-stable region with multiple classical solutions (M) or a region with a unique but parametrically unstable solution (P), see \figref{fig:JPD1}(c), Ref.~\cite{Sarchi2008} and supplementary material.
%This parametric instability (P) can be used for strong amplification of small fluctuations around the classical steady state.
%Close to the phase transition from the stable into one of the unstable regions the presence of quantum fluctuations is important. At this bias point, small signals around the classical steady state are amplified.
%We distinguish two cases of special relevance.
When driving the dimer close to either one of the two red-shifted eigenfrequencies $\tilde{\omega}_-$ or $\tilde{\omega}_+$, indicated by the blue dashed lines in  \figref{fig:JPD1}(c), the system behaves like a single nonlinear cavity leaving the undriven mode idle. Close to the phase transition into the multi-stable region (M) the finite on-site interaction strength $U$ stimulates the generation of signal-idler photon pairs. Since both signal and idler fields occupy the same mode, this is a degenerate parametric amplification process \cite{Castellanos2008}, see schematic representation in \figref{fig:JPD1}(d).
%This results in squeezing and phase-sensitive amplification  near the phase transition for which signal and idler fields occupy the same mode, see \figref{fig:JPD1}(d).
However, when driving the dimer in between the two eigenfrequencies $\tilde{\omega}_-$ and $\tilde{\omega}_+$ (red line in \figref{fig:JPD1}(b)), resulting in an equal population of the symmetric and antisymmetric mode, we observe a fundamentally different behavior. In this case we approach the transition from the stable (S) into the parametrically instable region (P), near which quantum fluctuations stimulate the generation of entangled photon pairs into the symmetric and anti-symmetric mode at a rate diverging at the phase transition. When additional signal fields are applied to the dimer, nondegenerate amplification is expected at a large detuning on the order of $2J$ between signal and idler modes (\figref{fig:JPD1}(e)).
%Due to the generality of the underlying Bose-Hubbard-dimer Hamiltonian the implementation of the presented scheme is not limited to superconducting circuits only but may also be realized with mechanical, %\cite{Teufel2011,Okamoto2013,Karabalin2009},
%optical or atomic systems.

\begin{figure}[b]
\centering
\includegraphics[scale=1]{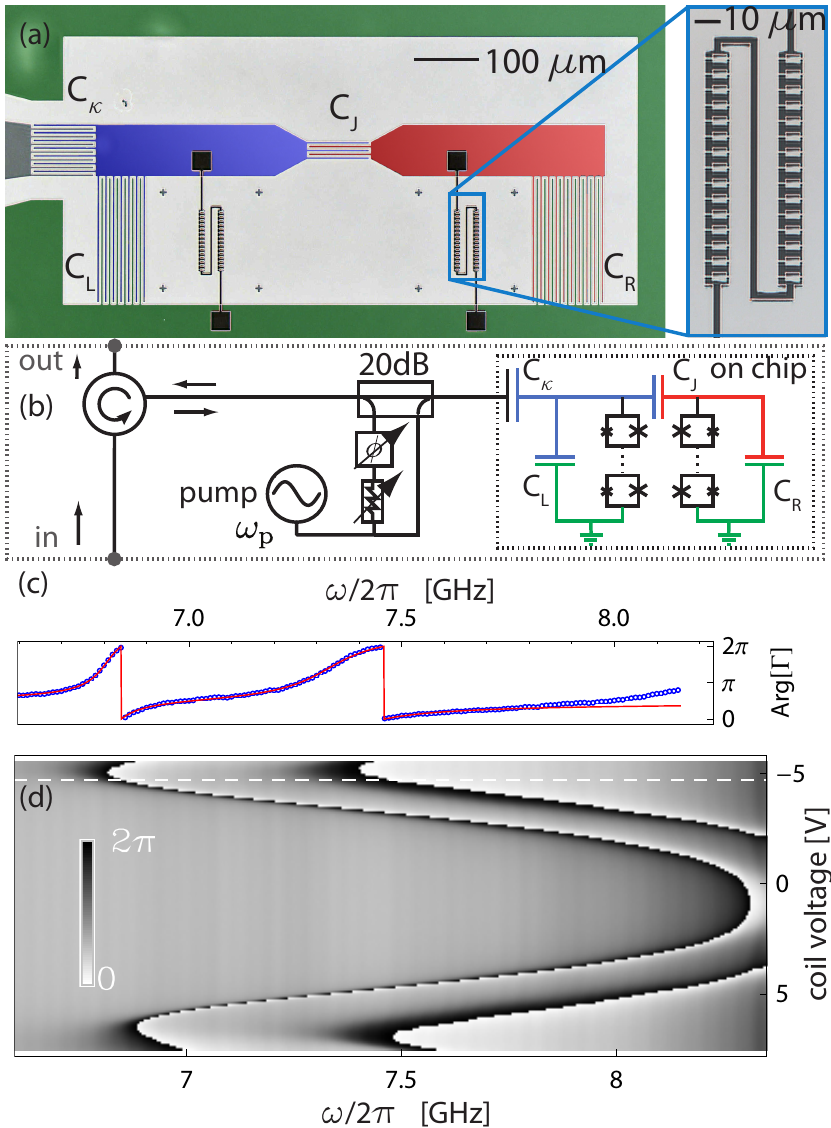}
\caption{(a) False-color micrograph of the sample. The inter-digitated finger structures form the capacitors of two coupled oscillators. An effective nonlinear inductance is realized as an array of SQUIDs in each resonator, also shown enlarged. (b) Circuit diagram of the experimental setup. The circuit is driven with a pump field at frequency $\omega_p = \omega_0+\delta$ through a -$20\,{\rm dB}$ directional coupler, of which the second port is used to interferometrically suppress the pump field reflected from the sample by more than ${\text -}60\,$~dB. Input and output signal fields are separated using a circulator. (c) Argument Arg$[\Gamma]$ of the measured (blue dots) and fitted (red line) reflection coefficient $\Gamma$  {\it vs}. probe frequency. (d) Measured Arg$[\Gamma]$ {\it vs}.  external magnetic flux. The white dashed line indicates the data trace shown in (c).}
\label{fig:JPD2}
\end{figure}

We demonstrate the phenomena discussed above in a circuit QED implementation of the Bose-Hubbard dimer which \CEE{we chose to refer to as a} Josephson parametric dimer (JPD).  In the JPD two interdigitated finger capacitors $C_L$ and $C_R$ shunted by an array of superconducting quantum interference devices (SQUIDs) form two lumped element LC oscillators in which the SQUIDs act as inductors (\figref{fig:JPD2}(a)). The SQUID inductance and with that the resonance frequencies of the JPD circuit is tuned by applying an external magnetic field through a coil mounted on the sample holder. The SQUID nonlinearity leads to effective photon-photon interactions with a strength \cite{Eichler2014} $U/2\pi \approx$ -$E_c/h M^2\approx$ -$80\,{\rm kHz}$, which depends on the charging energy $E_c\approx e^2/2C_R$ and can be controlled by varying the number of SQUIDs $M$ in the array. To keep the effect of inhomogeneities in the array at a minimum we set a lower bound to the critical current of each SQUID by using junctions with a ratio of Josephson energies given by $E_{J,1}/E_{J,2}\approx 1/3$. The coupling rate between the two resonators $J\approx C_J \omega_0/4 C_R $ is proportional to the capacitance $C_J$. $C_\kappa$ determines the coupling $\kappa$ to the in and output line.
%. The environmental coupling is $\kappa\approx C_\kappa^2 \omega_L^2 Z_0/(C_\kappa+C_L)$, where $Z_0=50\,\Omega$ is the transmission line impedance.
%While the oscillator on the right hand side is decoupled from the environment, the resonator on the left hand side is coupled with rate $\kappa$ to the transmission line through the capacitor $C_\kappa$.
A circuit diagram of the JPD device illustrating its operation as an amplifier is shown in \figref{fig:JPD2}(b).
%Further decoupling between the input and output of the amplifier is achieved by additional isolators (not shown).

\begin{figure*}[t]
\centering
\includegraphics[scale=0.95]{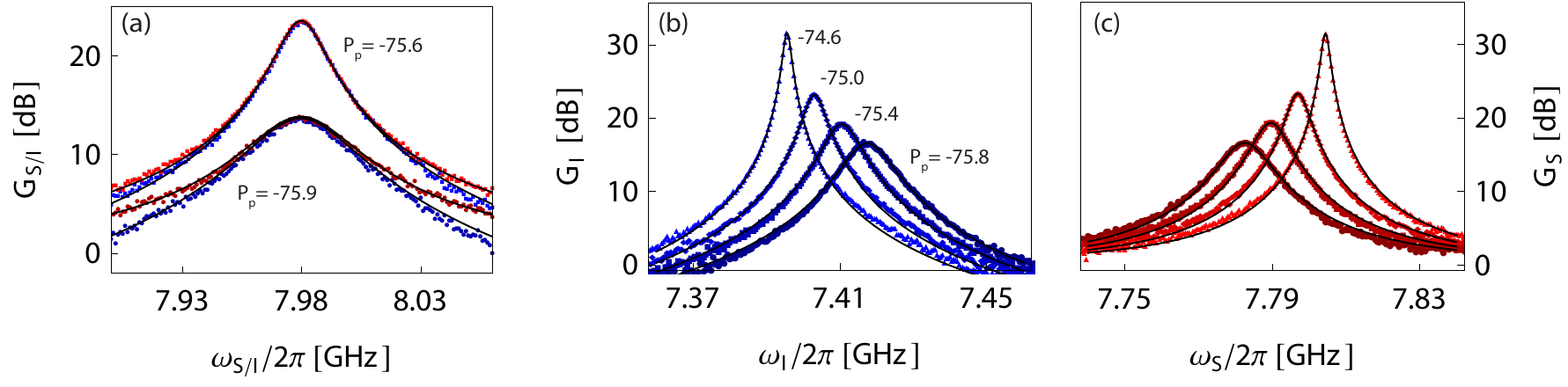}
\caption{(a) Measured signal (red) and idler gain (blue) {\it vs.} frequency for two pump powers $P_p \approx \{-75.9, -75.6\}\,$dBm fitted to a Lorentzian (black lines) for degenerate operation. (b,c) Measured idler $G_I$ and signal gain $G_S$ for nondegenerate operation together with Lorentzian fits (black lines) for drive powers $P_p \approx \{-75.8, -75.4, -75., -74.6\}\,$dBm.
}
\label{fig:JPD3}
\end{figure*}
We have measured the argument of the reflection coefficient ${\rm Arg}[\Gamma]$ of the JPD in linear response using a weak test tone of frequency $\omega/2\pi$. As expected, we find two resonances each leading to a phase shift of $2\pi$ in the reflected signal (\figref{fig:JPD2}(c)). By fitting the data to the model obtained from input-output theory (supplementary material), we extract the parameters $(\omega_L,\omega_R,\kappa,J)/2\pi\approx(7.0,7.2,0.29,0.25)\,{\rm GHz}$ for this bias point.  The left and right modes are found to be sufficiently close to resonance $|\omega_L-\omega_R| \lesssim \kappa$ as desired. By varying the external magnetic field through the SQUID arrays we tune both modes simultaneously (\figref{fig:JPD2}(d)).

To achieve degenerate amplification we drive the JPD with a coherent pump tone at frequency $\omega_p/2\pi = 7.98\,$GHz close to the resonance frequency of the symmetric mode $\tilde{\omega}_+$.  As expected, the measured signal and idler gain (red and blue points) is to very good approximation described by Lorentzian lines (black solid lines) for the two indicated pump powers (\figref{fig:JPD3}(a)).

While the degenerate amplification process is the conventional one in Josephson parametric amplifiers \cite{Castellanos2008}, we observe nondegenerate parametric amplification when driving the JPD in between the symmetric and anti-symmetric mode. In this case the symmetric and anti-symmetric modes of the JPD serve as signal and idler modes (Figs.~\ref{fig:JPD3}(b,c)). When amplifying a signal at $\omega_s/2\pi=7.79\,$GHz (\figref{fig:JPD3}(c)), the idler field is far detuned from the signal field at $\omega_i/2\pi=(2\omega_p-\omega_s)/2\pi = 7.41\,$GHz (\figref{fig:JPD3}(b)) allowing for simple rejection from the detection band for typical bandwidths of less than $J$ as required for phase preserving amplification.  For the chosen drive parameters the gain curves are well described by Lorentzian lines (black lines).
%The full expression for the frequency dependent gain (see supplementary material) suggests that non-Lorentzian gain profiles with bandwidths not limited by $\sim\kappa/\sqrt{G}$ can be achieved. In addition we suggest that as an alternative to the designs presented in Refs.~\cite{HoEom2012, Mutus2014} parametric amplifiers with well-controllable broadband amplification can be realized by extending the presented design to multi-cavity arrays.
%, which do not rely on hardly controllable environmental effects.
In contrast to other implementations of nondegenerate parametric amplifiers \cite{Bergeal2010}, the JPD emits the signal and the idler fields into the same transmission line, which  can therefore both be used for amplification. Furthermore,
%we unify .... its
our compact lumped element design and the use of SQUID arrays improves the achievable bandwidth and the dynamic range \cite{Eichler2014} compared to existing nondegenerate amplifiers while keeping a wide tunability.

For the measurements shown in  \figref{fig:JPD3}(c) we have controlled the gain by varying the pump power. The fact that we can reach gain values of more than $ 50\,{\rm dB}$ (not shown) indicates that the amplifier is far from saturation when operated at moderate gain. In fact, for samples with larger $\kappa/|U|$, we have measured amplification with remarkable dynamic range specified by a $1\,$dB compression point at input signals of $-110\,$dBm ($2000$ photons per $\mu$s) at a gain of $20\,$dB, as well as gain-bandwidth-products of more than $250\,$MHz in the nondegenerate mode of operation. Further improvements in the dynamic range are expected to be straightforward to realize by increasing the number of  SQUIDs in the arrays, by using Josephson junctions with even larger Josephson energy and by increasing $\kappa$.

In our experiments we probe the quantum nature of the observed parametric conversion processes by leaving the JPD input in the vacuum state and observing the creation of entangled photon pairs.
%To demonstrate quantum limited performance of the JPD and its use as a source of entangled photons, we have measured correlations between the signal and idler fields.
%in the absence of applied signal fields
The two-mode squeezing spectrum $S_{+-}^\phi(\Delta) \propto {\rm Var}[ e^{-i \phi}a_{\Delta} +e^{ i \phi}a^\dagger_{-\Delta}]$ (supplementary material) is a direct measure of this Einstein-Podolsky-Rosen (EPR) type entanglement \cite{Miranowicz2010} and allows us to resolve the asymmetric frequency-dependence of squeezing correlations. Here, $a_{\Delta}$ ($a_{-\Delta}$) is the annihilation operator for signal (idler) photons at detuning $\Delta$ ($-\Delta$) from the pump and $\phi$ is the phase of the pump relative to the local oscillator used for detection.
%We measure the squeezing spectrum by down-converting the emitted radiation at a microwave mixer with a local oscillator field resonant with the pump field $\omega_{\rm LO}=\omega_{p}$.
\begin{figure}[b]
\centering
\includegraphics[scale=1]{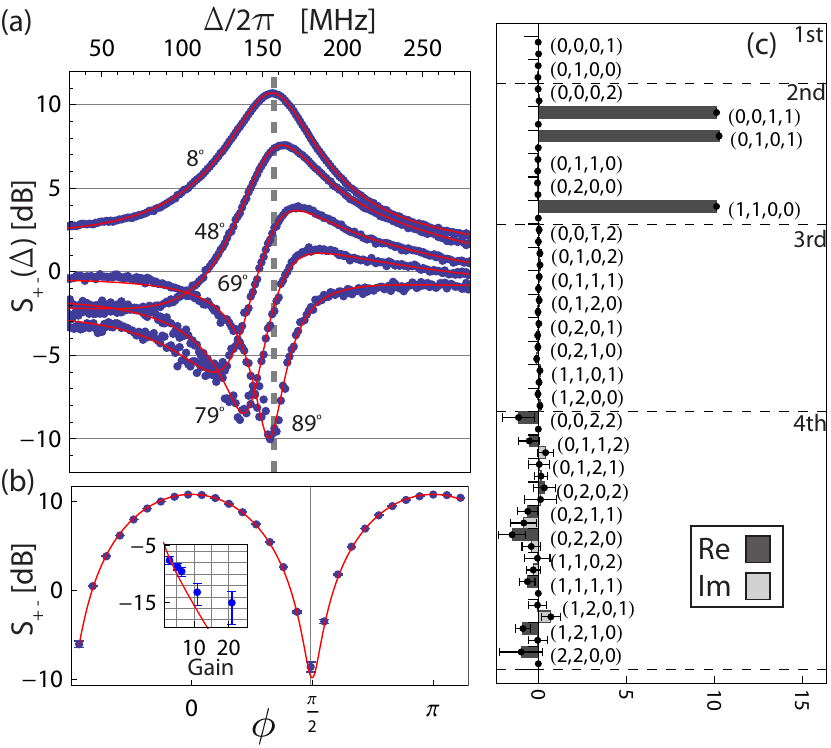}
\caption{(a) Two-mode squeezing spectrum $S_{+-}^{\phi}(\Delta)$ for local oscillator phases $\phi=\pi\{8, 48, 69, 79, 89\}/180$ with a global fit to the theoretical model. (b) Value of $S_{+-}^{\phi}(\Delta)$ at the sideband frequency $\Delta/2\pi=157.5\,$MHz indicated by the vertical dashed line in (a) \textsl{vs.} local oscillator phase $\phi$. The inset shows the value of squeezing at $\phi=\pi/2$ \textsl{vs.} gain in comparison to the ideal theoretical value (red line). (c) Real and imaginary part of the measured cumulants $\langle\langle(a_+^\dagger)^{n}(a_+)^m (a_-^\dagger)^{k}(a_-)^l\rangle\rangle$ for indicated orders $(n,m,k,l)$ up to $n+m+k+l\leq4$.}
\label{fig:JPD4}
\end{figure}
%By sampling and Fourier transforming the first order correlation function of the down-converted signal we measure the noise spectrum as a function of sideband frequency $\Delta$. These measurements have been realized using a \emph{Virtex 6} FPGA from \emph{Xilinx} enabling a measurement bandwidth of 500 MHz.
Depending on the phase $\phi$ we observe noise squeezing (anti-squeezing) below (above) the vacuum limit. The measured squeezing spectra fit very well to our theoretical model (supplementary material) and accurately reproduce the spectral asymmetry for intermediate LO phases shown on a logarithmic scale relative to the vacuum level ($0\,$dB) in \figref{fig:JPD4}(a). The spectra also demonstrate that measured squeezing and anti-squeezing are almost perfectly inversely proportional to each other. The value of the squeezing spectrum evaluated at the detuning indicated by the vertical dashed line in \figref{fig:JPD4}(a) shows the expected sinusoidal dependence on the phase $\phi$, see fit to theory (red line) in \figref{fig:JPD4}(b). The gain-dependent squeezing reaches values down to below $-12\,$dB (see inset) in a bandwidth larger than $10\,{\rm MHz}$ which is, to the best of our knowledge, the largest value reported so far for superconducting circuits \cite{Castellanos2008,Eichler2011a,Flurin2012,Menzel2012}.

\CEEE{To further investigate the statistical properties of the signal $a_+$ and idler $a_-$ fields and their correlations we have measured cumulants $\langle\langle(a_+^\dagger)^{n}(a_+)^m (a_-^\dagger)^{k}(a_-)^l\rangle\rangle$ with order $(n,m,k,l)$ up to $n+m+k+l\leq4$. The cumulant representation of correlators is particularly suitable to determine how well the analyzed radiation fields are described by ideal Gaussian states, since only the quadratic terms are expected to be non-zero.
%For this experiment we have used two detection channels to individually record the signal and idler radiation each in a $4\,$MHz band around their carrier frequencies \cite{Eichler2011a,Lang2013}.
Based on histograms of the measured field quadratures (supplementary material) we have extracted the cumulants. While the $(1,1,0,0)$ and the $(0,0,1,1)$ terms describe the average quadrature fluctuations in the signal and idler fields, respectively, the large (0,1,0,1) term demonstrates the entanglement correlations between the two fields (\figref{fig:JPD4}(c)). Except for the second order terms all higher order cumulants vanish, as expected for an ideal Gaussian state. The Gaussian property is an essential requirement, when using this signal-idler entanglement as a resource in continuous variable quantum computation protocols \cite{Braunstein2005}. This property is also highly relevant when employing the JPD for photon correlation measurements in which the statistical properties of the amplified field are to be preserved. Our measurements highlight the excellent performance of the presented device and its potential to be broadly used in cryogenic setups aiming at quantum limited measurements -- particularly in superconducting circuits.
}
\\\\
We acknowledge helpful discussions with Alexandre Blais. This work was supported by the European Research Council (ERC) through a Starting Grant, by the NCCR QSIT and by ETHZ.



\section*{Supplementary material}

# Supplementary material for "Quantum limited amplification and entanglement in coupled nonlinear resonators"

C. Eichler[1], Y. Salathe[1], J. Mlynek[1], S. Schmidt[2], A. Wallraff[1]

[1]*Department of Physics, ETH Zürich, CH-8093, Zürich, Switzerland. and*

[2]*Institute for Theoretical Physics, ETH Zürich, CH-8093, Zürich, Switzerland.*

(Dated: April 8, 2014)

## I. THEORY

In this section we derive input-output relations for a driven dissipative Bose-Hubbard dimer described in the laboratory frame by the Hamiltonian

$$H/\hbar = \omega_L a_L^\dagger a_L + \omega_R a_R^\dagger a_R + J(a_L a_R^\dagger + a_R a_L^\dagger) + \frac{1}{2}(U_L (a_L^\dagger)^2 a_L^2 + U_R (a_R^\dagger)^2 a_R^2). \qquad (1)$$

While mode $a_R$ is subject to internal loss at rate $\gamma$, mode $a_L$ is coupled with rate $\kappa \gg \gamma$ to the continuum of modes in a one-dimensional transmission line. The Langevin equations for this model read

$$\begin{aligned} \dot{a}_L &= -i(\omega_L - i\kappa/2) a_L - iJ a_R - i U_L a_L^\dagger a_L^2 + \sqrt{\kappa} a_{in}. \\ \dot{a}_R &= -i(\omega_R - i\gamma/2) a_R - iJ a_L - i U_R a_R^\dagger a_R^2. \end{aligned} \qquad (2)$$

In the following, we decompose all fields into a classical coherent part $\alpha$ and quantum fluctuations $b$, i.e.,

$$a_i \equiv (\alpha_i + b_i(t)) e^{-i\omega_p t}. \qquad (3)$$

By multiplying with $e^{-i\omega_p t}$ we chose to work in a frame rotating at the pump frequency $\omega_p$. We also assume that the input is composed of a large coherent tone $\alpha_{\text{in}}$ and a small quantum signal $b_{\text{in}}(t)$ with $\langle b_{\text{in}}^\dagger(t) b_{\text{in}}(t)\rangle \ll |\alpha_{\text{in}}|^2$ such that we can neglect quadratic and higher order quantum fluctuations in Eq. (2). This leads to a closed set of nonlinear equations of motion for the classical fields,

$$\begin{aligned} \dot{\alpha}_L &\equiv 0 = i\left(\delta_L + i\kappa/2\right)\alpha_L - iJ\alpha_R - iU_L|\alpha_L|^2 \alpha_L + \sqrt{\kappa}\alpha_{\text{in}} \\ \dot{\alpha}_R &\equiv 0 = i\left(\delta_R + i\gamma/2\right)\alpha_R - iJ\alpha_L - iU_R|\alpha_R|^2 \alpha_R \end{aligned} \qquad (4)$$

with $\delta_i = \omega_p - \omega_i$ and a linear system of equations for the fluctuations

$$\dot{\mathbf{b}}(t) = \mathcal{S}\mathbf{b}(t) + \sqrt{\kappa}\mathbf{b}_{\text{in}}(t). \qquad (5)$$

Here, we have defined $\mathbf{b} = (b_L, b_L^\dagger, b_R, b_R^\dagger)^\mathrm{T}$, $\mathbf{b}_\mathrm{in} \overset{\gamma \ll \kappa}{\approx} (b_\mathrm{in}, b_\mathrm{in}^\dagger, 0, 0)^\mathrm{T}$ and the matrix

$$\mathcal{S} = \begin{pmatrix} -i\tilde{\delta}_L & -iU_L\alpha_L^2 & -iJ & 0 \\ iU_L\alpha_L^{*2} & i\tilde{\delta}_L & 0 & iJ \\ -iJ & 0 & -i\tilde{\delta}_R & -iU_R\alpha_R^2 \\ 0 & iJ & iU_R\alpha_R^{*2} & i\tilde{\delta}_R \end{pmatrix} \qquad (6)$$

with $\tilde{\delta}_L = -\delta_L - i\kappa/2 + 2U_L|\alpha_L|^2$ and a similar definition for $\tilde{\delta}_R$ (with $\kappa$ replaced by $\gamma$). The output field in the transmission line is then obtained from the solution of Eqs. (4) and (5) using the input-output relation [1]

$$a_\mathrm{out} = \sqrt{\kappa} a_L - a_\mathrm{in} \qquad (7)$$

where $a_\mathrm{out}$ is again decomposed into a classical and a quantum part as described by Eq. (3).

### A. Reflection coefficient for weak probe fields

In the absence of an input signal ($b_\mathrm{in} = 0$) and in the quasi-linear regime with negligible nonlinearity $U_{L(R)}|\alpha_\mathrm{in}|^2 \ll \kappa^2$, we solve Eq. (4) for the steady state explicitly and obtain from the input-output relation in Eq. (7) an analytic expression for the reflection coefficient

$$\Gamma \equiv \frac{\alpha_\mathrm{out}}{\alpha_\mathrm{in}} = -1 + \frac{\kappa(\frac{\gamma}{2} - i(\omega_p - \omega_R))}{J^2 - (i\frac{\kappa}{2} + \omega_p - \omega_L)(i\frac{\gamma}{2} + \omega_p - \omega_R)}. \qquad (8)$$

This expression has been used to fit the data shown in Fig. 2(c,d) of the main text.

### B. Steady state and stability

When solving for the steady state of Eq. (4) one obtains, in general, multiple classical solutions. The stability of each of these solutions is determined by the eigenvalues $\epsilon_\alpha$ of the (stability) matrix (6). If the real part of all eigenvalues is negative, i.e., $\mathcal{R}[\epsilon_\alpha] < 0$, the linear fluctuations in Eq. (5) are damped and the classical steady state solution is stable. However, if at least one of the real parts of the eigenvalues becomes positive, the steady state solution is unstable with respect to small fluctuations. In the phase diagram in Fig. 1 of the main text we have plotted three different regions corresponding to the existence of a unique and stable solution (S), a unique but parametrically unstable solution (P), and multiple classical solutions (M). For moderate drive strength, a large part of the region with multiple solutions (M) corresponds to bistable behaviour, which arises when the pump frequency is near resonant with the (red-shifted) symmetric or antisymmetric modes of the Bose-Hubbard dimer. If the drive strength is slightly below such a bistability, the

JPD can be used as a degenerate parametric amplifier. In the parametric unstable region (P), the pump can efficiently create signal and idler photons with large energy difference on the order of the hopping strength $2J$. In this regime, the JPD thus operates as a non-degenerate amplifier. A more detailed discussion of the phase diagram of the driven, dissipative Bose-Hubbard dimer is presented in Ref. [2].

### C. Amplifier gain

We now calculate the gain for the frequency components of the output signal relative to a small input field $b_{\text{in}}(t)$. For this purpose, we introduce the Fourier transform

$$b_i(t) = \int_{-\infty}^{\infty} d\Delta/(\sqrt{2\pi})\, b_i(\Delta)\, e^{-i\Delta t}, \qquad (9)$$

where $\Delta$ denotes the detuning of the signal from the pump frequency. With Eq. (9) we obtain from Eq. (5) the general solution $\mathbf{b}(\Delta) = \mathcal{G}(\Delta)\mathbf{b}_{\text{in}}(\Delta)$. Here, the gain matrix is given by

$$\mathcal{G}(\Delta) = (-i\Delta\mathcal{I} - \mathcal{S})^{-1}, \qquad (10)$$

such that the quantum part of the output field in Eq. (7) can be written as

$$b_{\text{out}}(\Delta) = g_S(\Delta)\, b_{\text{in}}(\Delta) + g_I(\Delta)\, b_{\text{in}}^\dagger(-\Delta), \qquad (11)$$

with complex signal

$$g_S(\Delta) = \kappa\mathcal{G}_{11}(\Delta) - 1 \qquad (12)$$

and idler gain coefficients

$$g_I(\Delta) = \kappa\mathcal{G}_{12}(\Delta). \qquad (13)$$

Note that the inverse in Eq. (10) can be calculated analytically resulting in explicit but rather lengthy algebraic expressions for signal and idler gain, which we omit here for brevity. In Fig. S1(a) we plot the intra-cavity fields $\alpha_L$ and $\alpha_R$ for a drive detuning $\delta = \delta_L = \delta_R = 0.2\,\kappa$. The left and right cavities have almost equal pump amplitude resulting in degenerate parametric amplification S1(b). For the detuning $\delta = -0.5\,\kappa$ and appropriate drive amplitude the pump photons appear predominantly in the right cavity S1(c) leaving the left cavity almost empty. In this case we get two gain peaks around the symmetric and anti-symmetric eigenfrequencies as desired for non-degenerate amplification (S1(d)).

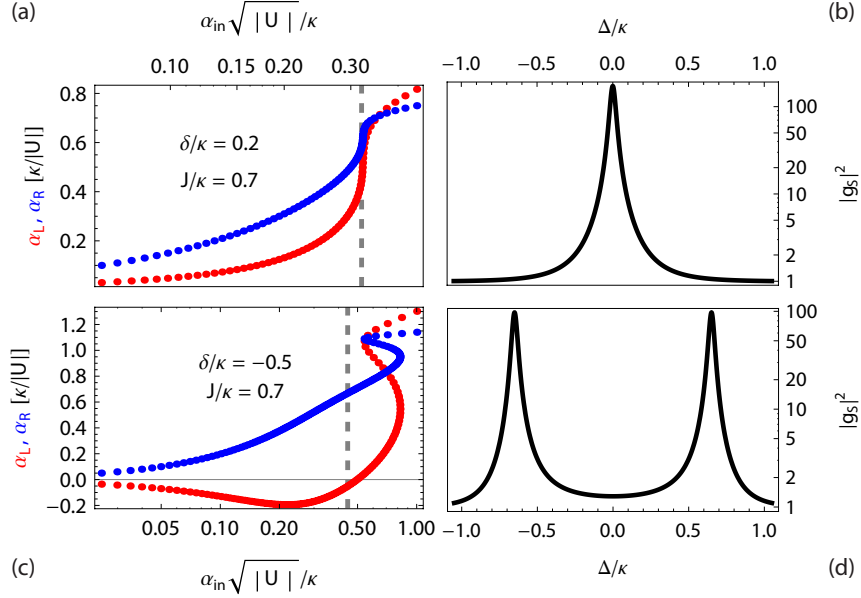

FIG. S1: (a,c) Calculated steady state intra-cavity fields $\alpha_L$ (red) and $\alpha_R$ (blue) as a function of drive amplitude for the specified drive detuning $\delta$ and hopping rate $J$. All quantities are in scale-invariant units. (b,d) Power gain $|g_S(\Delta)|^2$ vs. signal detuning $\Delta$ for drive amplitudes indicated by the vertical dashed line in corresponding panels (a) and (c), respectively.

### D. Squeezing spectrum

The squeezing spectrum $S^\phi_{+-}(\Delta)$ in units of photons per Hz per sec is defined as

$$S^\phi_{+-}(\Delta)\delta(\Delta - \Delta') \equiv \langle X^{\phi\,\dagger}_{\Delta'} X^\phi_\Delta \rangle, \qquad (14)$$

with the generalized quadrature $X^\phi_\Delta = e^{-i\phi} b_{\text{out}}(\Delta) + e^{i\phi} b^\dagger_{\text{out}}(-\Delta)$ and the Dirac distribution $\delta(\Delta - \Delta')$. Experimentally we measure the quantity $X^\phi_\Delta$ by mixing the output radiation with a local oscillator field at frequency $\omega_{\text{LO}} = \omega_p$ and phase $\phi$ relative to the pump field, and by sampling the voltage at the output of the mixer. Using the input output relation (11) and taking $\gamma = 0$ we obtain

$$S^\phi_{+-}(\Delta) = 1 + 2|g_I(\Delta)|^2 + 2\text{Re}[e^{-2i\phi} g_S(-\Delta) g_I(\Delta)] \qquad (15)$$

when vacuum noise is incident to the JPD obeying the correlation $\langle b_{\text{in}}(\Delta) b^\dagger_{\text{in}}(\Delta') \rangle = \delta(\Delta - \Delta')$. Additional thermal fluctuations with approximately frequency independent mean photon number $n_{\text{th}}$ result in a scaling of this expression by the factor $1 + 2n_{\text{th}}$. For appropriate relative phase $\phi$ the squeezing spectrum becomes smaller than the vacuum noise level at $0\,\text{dB}$, indicating the entanglement between signal and idler photons generated at positive and negative detuning from the pump.

## II. EXPERIMENTAL DETAILS

### A. Sample parameters and fabrication

The capacitive structures of the sample shown in Fig. 2(a) of the main text are fabricated from a niobium thin film sputtered onto a sapphire wafer and using a reactive ion etch defined by photolithography. From finite element simulations of the static electrical field we find lumped element capacitances $(C_L, C_R, C_\kappa, C_J) = (192, 257, 78, 29)\,\text{fF}$. The Josephson junctions are fabricated in a second step using electron beam lithography and double-angle shadow evaporation of aluminium. The overlap area of the small (big) junction of each SQUID is designed to be $1.1\,\mu\text{m}^2$ ($3.4\,\mu\text{m}^2$). The normal state resistance measurement of the fabricated Josephson junction array corresponds to a Josephson energy $E_J/h \approx 4150\,\text{THz}$ per SQUID, which is equivalent to a total Josephson inductance of $L_J \approx 1.2\,\text{nH}$ for the array with 30 SQUIDs. Additional geometric inductance is expected to be almost an order of magnitude smaller than the Josephson inductance. The smooth dependence of resonance frequencies on the applied magnetic field (Fig. 2(d)) indicates a homogeneous coupling to the external magnetic field, which we attribute to the large non-superconducting area around the SQUID loops and the use of asymmetric SQUIDs in the array.

### B. Dynamic range and bandwidth

We have also fabricated and characterized samples with larger coupling rates $\kappa$ and smaller nonlinearity $U$, for which we obtain larger bandwidth and dynamic range. The values stated in the main manuscript are based on the measurement data shown in Fig. S2. From the measured gain *vs.* signal frequency we extract a gain bandwidth product of $275\,\text{MHz}$, which is defined as the

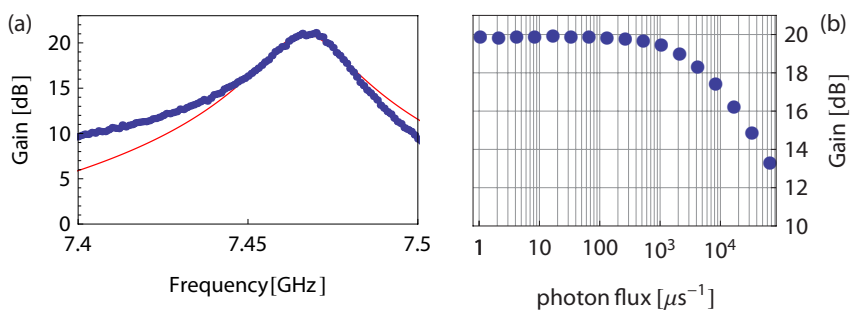

FIG. S2: (a) Measured gain *vs.* signal frequency fitted to a Lorentzian. (b) Measured gain *vs.* input signal power in units of photons per $\mu$s.

square root of the gain times the full width half maximum of the gain curve. From the data shown in panel (b) we find a 1 dB gain compression point at input signals with 2000 photons per $\mu$s.

### C. Correlation function and tomographic measurements

The frequency resolved squeezing spectrum is measured using an analog to digital (ADC) converter with 1 GSample/s interfaced with a field programmable gate array (FPGA). The radiation is down-converted with a local oscillator field using an IQ mixer. The voltage at the mixer output is then sampled with a 1 ns resolution. Individual data traces comprising 8192 samples are Fourier transformed in real time and multiplied with their complex-conjugate to obtain the noise spectrum of the sampled quantity. Averaging over $2^{23} \approx 8 \times 10^6$ repetitions results in the averaged spectrum.

To measure higher order cumulants we have performed a tomographic experiment similar to the one described in Ref. [3]. We individually detect the signal and idler radiation in a 4 MHz band around their carrier frequencies by using two heterodyne detection channels. In each channel we record both conjugate field quadratures $X$ and $P$, which are stored in a 4-dimensional histogram with axes $\{X_1, P_1, X_2, P_2\}$. Based on two histograms – one with the pump field turned on and one with the pump turned off – we evaluate the statistical moments $\langle (a_+^\dagger)^n (a_+)^m (a_-^\dagger)^k (a_-)^l \rangle$ up to order $n + m + k + l = 4$, see Ref. [4] for details. Here, $a_-$ ($a_+$) is the annihilation operator for a photon in the detected signal (idler) band. The normally ordered moments $\langle ... \rangle$ are transformed into the corresponding cumulants $\langle\langle ... \rangle\rangle$ based on the following identity

$$\langle\langle (a_+^\dagger)^n (a_+)^m (a_-^\dagger)^k (a_-)^l \rangle\rangle = \partial_x^n \partial_y^m \partial_z^k \partial_u^l \ln \sum_{a,b,c,d} \frac{\langle (a_+^\dagger)^a (a_+)^b (a_-^\dagger)^c (a_-)^d \rangle x^a y^b z^c u^d}{a!b!c!d!} \bigg|_{x=y=z=u=0},$$

where $\partial_y^m$ is the m-th partial derivative with respect to $y$ and ln is the natural logarithm. Error bars in Fig. 4(c) are extracted from the standard deviation of repeated measurements.

---


\end{document}